\documentclass[10pt,journal]{IEEEtran}
\usepackage{subfig}
\usepackage{amsmath}
\usepackage{graphicx}
\usepackage{xcolor}
\usepackage[noadjust]{cite}
\graphicspath{{figures/drawings/}{figures/plots/pdf/}}

\begin{document}

\title{Destination-Feedback Free Distributed Transmit Beamforming using Guided Directionality}

\author{Samer~Hanna,~\IEEEmembership{Student~Member,~IEEE,}
	Enes~Krijestorac,~\IEEEmembership{Student~Member,~IEEE,}
	and Danijela~Cabric,~\IEEEmembership{Fellow,~IEEE}%
	\thanks{The authors are with the Electrical and Computer Engineering Department, University of California, Los Angeles, CA 90095, USA. 	e-mails: 	\mbox{samerhanna@ucla.edu}, enesk@g.ucla.edu, danijela@ee.ucla.edu}
	\thanks{This work was supported in part by the CONIX Research Center, one of six centers in JUMP, a Semiconductor Research Corporation (SRC) program sponsored by DARPA.}}
{}

\maketitle

\begin{abstract}
Distributed transmit beamforming enables cooperative radios to act as one virtual antenna array, extending their communications' range beyond the capabilities of a single radio. Most existing  distributed beamforming approaches rely on  the destination radio sending feedback to adjust the transmitters' signals for coherent combining. However, relying on the  destination radio's feedback limits the communications range to that of a single radio. Existing destination-feedback-free approaches rely on phase synchronization and knowing the node locations with sub-wavelength   accuracy, which becomes impractical for radios mounted on high-mobility  platforms like  UAVs.  In this work, we propose and demonstrate a destination-feedback-free  distributed beamforming approach that leverages the radio's mobility and coarse location information in a dominant line-of-sight channel.  In the proposed approach,  one radio acts as a guide and moves to point the beam of the remaining radios towards the destination. We specify the radios' position requirements  and verify their relation to  the combined signal at the destination using simulations.  A proof of concept demo was implemented  using software defined radios, showing up to 9dB SNR improvement  in the beamforming direction just by relying on the coarse  placement of four radios.
\end{abstract}


\IEEEpeerreviewmaketitle

\section{Introduction}
Distributed transmit beamforming (DBF) is a cooperative communications technology that enables a group of  radios to act as a virtual antenna array. In distributed transmit beamforming, a group of synchronized radios sending the same message adjust their signals to ensure coherent combining at the destination. For a group of $N$ radios, distributed beamforming provides $N^2$ increase in the received power~\cite{mudumbai_distributed_2009}.
This power increase can be used to extend the communications range or reduce the power transmitted from the radios. Both communications range and power efficiency are of great importance for  unmanned aerial vehicles (UAVs), which have a limited power budget and are advantageous to deploy in large numbers at remote areas~\cite{shakhatreh_unmanned_2019}.

To achieve coherent combining using  DBF, the radios need to  synchronize their carrier frequencies and their symbol timing as well as adjust their phases to ensure coherent combining at the destination~\cite{mudumbai_distributed_2009}. Synchronization needs to happen among the beamforming radios and it does not depend on the destination radio. The phase correction, however, depends on the destination. There are two methods to adjust the phases for distributed beamforming~\cite{jayaprakasam_distributed_2017,ochiai_collaborative_2005}: the first one (\emph{Feedback DBF}) relies on the destination radio assisting the nodes in obtaining  channel phase estimates, while the second method (\emph{Location DBF}) relies on the nodes knowing their locations and the beamforming direction. 

Feedback DBF assumes that the destination can communicate with the beamforming radios. This communication can be in the form of a preamble transmitted from the destination~\cite{ochiai_collaborative_2005}, or the destination sending back the channel estimates~\cite{yung-szu_tu_coherent_2002}, or just providing binary feedback with the DBF radios randomly perturbing their phases~\cite{mudumbai_distributed_2010}.  Many of these methods were demonstrated using software defined radios~\cite{jayaprakasam_distributed_2017}. However, while these approaches can correct the phase to attain coherent combining, they rely on the destination radio having sufficient transmit power to reach the DBF radios which limits the communication range to that of the destination radio regardless of how many DBF radios are used.

As for Location DBF, it   does not need any destination feedback  and relies only on the nodes knowing their relative locations. Using this information and the direction towards the destination,  the radios can calculate the phases needed for beamforming~\cite{jayaprakasam_distributed_2017}. 
However, to have the full beamforming (BF) gains using this approach, location information accurate to a fraction of a wavelength is necessary, and the gains degrade rapidly due to localization errors~\cite{ochiai_collaborative_2005}. 
This requirement places stringent localization requirements and limits the applicability of this approach for high  mobility platforms like UAVs where typically only coarse location is available using  satellite navigation systems. Additionally, this approach assumes that the DBF nodes are aligned in phase which is not easy to realize using radios having independent oscillators. While these requirements can be realized using communication among the DBF radios, it requires a large bandwidth~\cite{ellison_multi-node_2021}.  Another approach (\emph{Random DBF})  avoids destination feedback  by relying on the randomness of the combining gain from unsynchronized radios along with repeating transmissions~\cite{sklivanitis_testbed_2013}. However, this approach is not scalable and has a low throughput.

In this paper, we propose \emph{Guided Beamforming} as an approach for cooperating mobile radios  to attain coherent combining at a distant destination radio unable  to provide feedback. To overcome the lack of feedback, the DBF radios, which are assumed to be in proximity of each other,  need to know the beamforming direction to the destination and have a LOS channel between them. These radios can be mounted on ground robots,  UAVs, or handheld as long as they can be coarsely positioned relying on satellite navigation for instance.   Guided  DBF relies on assigning one of the DBF radios as a guide and the rest as followers. The followers adjust their signals to ensure coherent combining at the guide. Since the guide is close to the followers, it can provide them with reliable feedback for DBF unlike the destination. Using radios' mobility and coarse localization, the followers cluster and the guide moves towards the desired DBF direction to point the combined signal at the desired DBF direction. We verify this concept using simulations and analyze the position requirements of the guide and followers along with its sensitivity to localization errors and non-LOS channel components. Then, we demonstrate this approach using software defined radios. Using 4 DBF radios we were able to attain more than 3 fold increase in the signal magnitude (9x increase in power received) towards the direction of interest. To the best of the authors' knowledge, this is the first demonstration of distributed beamforming  that achieves coherent combining  at the destination without any destination feedback and  without sacrificing throughput with repeated transmissions nor requiring a large bandwidth.
This approach can be used to extend the range of communications towards a distant destination radio unable to provide any feedback to the DBF radios. 
Our main contributions can be summarized as follows:
\begin{itemize}
	\item We proposed Guided  DBF as an approach to enable distributed beamforming towards a  destination unable to provide feedback, assuming a LOS channel between the DBF radios. Our proposed approach leverages the radio's mobility and coarse localization to achieve coherent combining at the destination. 
	\item Using simulations, we showed that the proposed approach can tolerate DBF radios location errors  within multiple wavelengths in contrast to location based beamforming, which requires location accuracy within a fraction of a wavelength.
	\item We compared Guided DBF with Feedback DBF simulated on the signal level. We showed that Guided DBF provides significantly higher BF gains than Feedback DBF as the distance between the BF radios and the destination increases, highlighting the advantage of Guided DBF in increasing the communication range.
	\item We experimentally demonstrated Guided DBF using software defined radios. An average BF gain of over 3x was achieved in the intended direction when using 4 beamforming radios leading to 9dB SNR improvement on the average. The combining gains measured in different directions were shown to follow the expected DBF pattern predicted by simulations. 
	
\end{itemize}

\section{Related Work}
As discussed earlier, existing DBF approaches either rely on destination feedback  or highly accurate knowledge of radio locations. 

\textbf{Destination Feedback:} Many existing works have relied on destination feedback for coherent combining~\cite{mudumbai_distributed_2009,jayaprakasam_distributed_2017}. A system using explicit channel feedback was proposed in~\cite{yung-szu_tu_coherent_2002} and demonstrated in~\cite{leak_distributed_2018,mohanti_airbeam_2019,alemdar_rfclock_2021,hanna_distributed_2021} for communications and in~\cite{mohanti_wifed_2021} for energy transfer.  To reduce the feedback overhead, a  1-bit feedback algorithm was developed~\cite{mudumbai_distributed_2010}. Using this approach, the nodes randomly perturbate their phase and the receiver provides binary feedback indicating whether the channel has improved or no in an iterative manner. This approach was used in several experimental evaluations of distributed beamforming for instance  \cite{quitin_demonstrating_2012,rahman_fully_2012} and was proposed for energy transfer in~\cite{lee_distributed_2021}. Joint location and beamforming optimization was considered using destination feedback in~\cite{george_multi-agent_2020,george_model-free_2020}.  Motion and communications energy was  optimized for mobile robots  in~\cite{muralidharan_energy_2018} using destination feedback along with channel predictions.
In~\cite{iii_time-slotted_2008}, a synchronization algorithm based on roundtrip message exchanges was developed.  Other works have proposed using channel reciprocity  for channel estimation~\cite{ochiai_collaborative_2005} and this approach was demonstrated in~\cite{peiffer_experimental_2016}. All these approaches rely on the destination having sufficient transmit power to provide reliable feedback, which is not  the case for remote destinations.

\textbf{Location Based Beamforming:} Other works have relied on the knowledge of the locations for the beamforming radios to adjust the phases. Some works have focused on either studying the beampattern of random placements of radios or optimizing the beampattern~\cite{jayaprakasam_distributed_2017}. In~\cite{ochiai_collaborative_2005}, the beampatterns obtained using uniform random deployments of transmitters within a disk area was considered. The effects of phase jitter and location estimation errors on the beampattern were studied. Other works have studied the beampattern of beamforming nodes following a Gaussian distribution~\cite{ahmed_collaborative_2009} or arbitrary distributions~\cite{huang_collaborative_2012}. Among the works that considered beampattern optimization, some have proposed using node selection or coefficient perturbation to create a null in a certain direction~\cite{kong_simultaneous_2020}, minimize the beamwidth~\cite{zarifi_distributed_2009}, or control the sidelobes~\cite{liang_jssa_2019,sun_energy_2019}. These works  are only theoretic and typically assume accurate localization and phase synchronization.   Location DBF was implemented  in~\cite{ellison_multi-node_2021} based on ranging using 3 DBF radios placed in a linear array. To attain accurate localization a large bandwidth (BW) of 12.5 MHz was used and signals were exchanged among each pair of radios. While this setup avoids using feedback, it requires a specific arrangement, large bandwidth, and pairwise signal exchange makes its overhead scale quadratically with the number of DBF radios.

Other works have proposed and demonstrated zero-destination-feedback beamforming~\cite{sklivanitis_testing_2011,sklivanitis_testbed_2013}, which works by sending multiple repetitions of the signal and using the fact that unsynchronized carriers occasionally combine constructively. While this approach avoids relying on destination feedback and is simple to implement,  it negatively affects the throughput as multiple transmissions of the same message is needed.

\renewcommand{\b}[1]{\boldsymbol{\mathrm{#1}}}
\renewcommand{\mp}[1]{\b{p}_{#1}}
\newcommand{\mpx}[1]{p^{x}_{#1}}
\newcommand{\mpy}[1]{p^{y}_{#1}}
\newcommand{\mpz}[1]{p^{z}_{#1}}

\newcommand{\mdt}[1]{\Delta t_{#1}}
\newcommand{\mdf}[1]{\Delta f_{#1}}

\newcommand{\mhm}[1]{|h_{#1}|}
\newcommand{\mhp}[1]{\angle h_{#1}}
\newcommand{\mhpg}[1]{\angle g_{#1}}

\newcommand{\mph}[1]{\angle \hat{h}_{#1}}
\newcommand{\mdth}[1]{\Delta \hat{t}_{#1}}
\newcommand{\mdfh}[1]{\Delta \hat{f}_{#1}}

\newcommand{\mw}[1]{w_{#1}}
\newcommand{\mwm}[1]{\kappa_{#1}}
\newcommand{\mwp}[1]{\theta_{#1}}
\newcommand{\mwt}[1]{\tau_{#1}}
\newcommand{\mwf}[1]{\psi_{#1}}
\section{Problem Statement}
\begin{figure}[t!]
	\centering
	\includegraphics[width=3.3in]{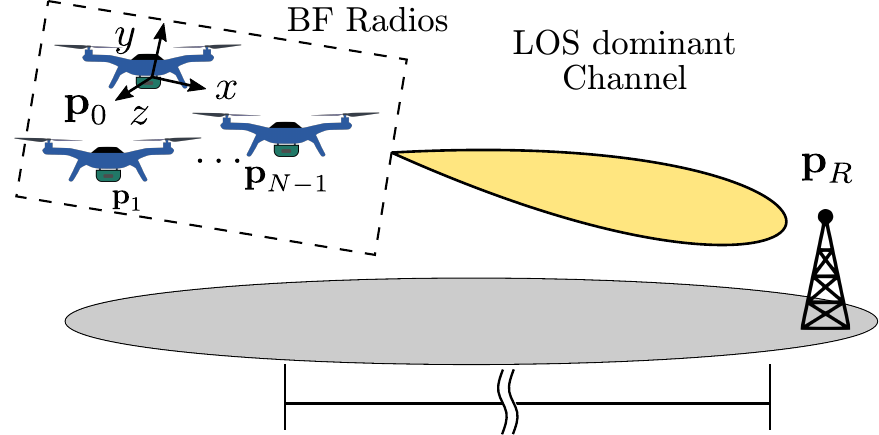}
	\caption{The objective of distributed beamforming is to coherently combine signal from $N$  radios, which are assumed to be mobile, at a distant destination radio.}
	\label{fig:system_model}
\end{figure}
Consider $N$ mobile radios that use DBF to send a critical message to  a   distant  destination radio   $R$. The DBF radios  are cooperating on the same task and hence are assumed to be close to each other and far from the destination radio $R$ beyond its communication range.  Hence, the destination cannot provide any feedback for DBF. The  DBF radios are assumed to know the beamforming direction toward the destination and their locations coarsely using a satellite navigation system like GPS.
For example, the radios can be mounted on UAVs performing a search and rescue operation in a remote area. They are communicating with the destination radio placed at  the operation center established at their takeoff location. Due to their elevation above the ground and assuming a deployment in a remote non-urban area, an air-to-ground channel is dominated by LOS propagation~\cite{khuwaja_survey_2018}. Hence, the knowledge of locations  is sufficient for the DBF radios to determine the DBF direction without feedback from the destination.%
 
  DBF radio $i$ is located at $\mp{i}=[\mpx{i},\mpy{i},\mpz{i}]^T$, where $\mpx{i}$, $\mpy{i}, \mpz{i}$ are the $x$, $y$, and $z$ coordinates of the node $i$, with respect to  node $0$ which is used as a reference, i.e, $\mp{0}=[0,0,0]^T$.  The destination  is located at $\mp{R}=[\mpx{R},\mpy{R},\mpz{R}]^T$ in the far field of the DBF radios such that $d_{i,R} >> d_{i,j}$ for all $i$ and $j$ from $0$ to $N-1$, where the distances are defined as $d_{i,R} = \|\mp{i}-\mp{R} \|$ and  $d_{i,j}=\|\mp{i} - \mp{j} \|$. Without a loss of generality, we assume that the known DBF direction is the positive x direction.  If a LOS channel exists towards the destination, the destination receiver would be located far on the $x$-axis  such that $|\mpx{R}|>>\sqrt{(\mpy{R})^2 +(\mpz{R})^2}$. This setup is shown in Fig.~\ref{fig:system_model}.

Assuming that the DBF radio are synchronized in time and frequency using an over-the-air synchronization protocol as discussed later in Section~\ref{sec:implementation}, the signal transmitted by radio $i$ is given by
\begin{equation}
x_{i}(t)= \Re\{ s(t) \mw{i} e^{j 2\pi  f_c t}\}
\end{equation}
where  $s(t)$ is the complex baseband payload containing the message, $f_c$ is the carrier frequency, and $\Re\{\cdot\}$ denotes the real part. The payload $s(t)$ is shared among all the BF radios  using a network broadcasting protocol~\cite{williams_comparison_2002}. To ensure coherent combining at the destination, each radio precodes its signal with a complex weight  $\mw{i}=\mwm{i}e^{j\mwp{i}}$, where $\mwm{i}$ is the magnitude and $\mwp{i}$ is the phase. The received signal at $R$ is given by
\begin{align}
y(t) &= \sum_{i=0}^{N-1} \Re\{ \mw{i} h_{i} e^{j2\pi f_c t} s(t) \}+ \gamma(t) \\
& = \sum_{i=0}^{N-1} \Re \{ \mwm{i} \mhm{i} e^{j (2\pi f_c t + \mwp{i}+\mhp{i})} s(t)\} + \gamma(t) 
\end{align}
where the narrowband channel is given by $h_{i}=\mhm{i}\exp(j\mhp{i})$,  $\mhm{i}$ is its magnitude,  $\mhp{i}$ is its phase,  and $\gamma(t)$ is the additive Gaussian noise.  %

The normalized magnitude of the beamforming gain is the ratio between the attained combining gain and the ideal combining gain and is given by
\begin{equation}
\Gamma = \frac{ |\sum_{i=0}^{N-1} \mwm{i} \mhm{i}    \cdot e^{j ( \mwp{i}+\mhp{i})}|}{ \sum_{i=0}^{N-1} \mwm{i} \mhm{i} }
\label{eq:DBF_gain}
\end{equation}
and it takes a value between 0 and 1. Perfect coherent combining at the destination occurs, if combining phases ($\mwp{i}+\mhp{i}$) are equal for all $i$, which this corresponds to $\Gamma=1$. A phase mismatch between the combining signals will lead to degraded BF gains.

 To maximize the power at the end receiver, each radio is assumed to transmit at its maximum power $\mwm{i}=\sqrt{P_T}$, which is optimal regardless of the channel magnitude~\cite{mudumbai_distributed_2009}.  Hence, our objective is to find the phases $\mwp{i}$ for coherent combining at the destination receiver. Thus for simplicity, we consider  normalized channel having $\mhm{i}$=1 for all $i$. 
  
As for modeling the  channel, we consider a Ricean channel  defined as follows
\begin{equation}
	h_{i} = \sqrt{\frac{K}{K+1}   } h^{\text{L}}_{i}+ \sqrt{\frac{1}{K+1} }   h^{N}_{i}
\end{equation}
where $h^{\text{L}}_{i}= e^{j \frac{2 \pi d_{i,R}}{\lambda}}$ models the geometric LOS channel component which depends on the positions of the radios with  $\lambda$ being the wavelength. As for the non-LOS channel component $ h^{N}_{i}$, it  accounts for the random reflections in the environment and it is modeled as a  standard complex Gaussian random variable. The value of $K$ determines the magnitude of the non-LOS component.  A $K$-factor equal to 0 corresponds to a rayleigh non-LOS channel and $K$-factor of infinity corresponds to a geometric LOS channel. For a LOS dominant channel  ($K \gg 1 $), the geometric LOS component is dominant and by changing the positions of the DBF radios ($\mp{i}$), we can change the channel with the destination and among the DBF radios.

Our objective is to determine the beamforming phases $\mwp{i}$  and optimize the  DBF radio positions $\mp{i}$ for $i\in \{0,\cdots, N-1\}$, to ensure coherent combining at the destination receiver (large~$\Gamma$). To enable DBF beyond the communications range of the destination, we do not rely on its assistance in calculating $\mwp{i}$. Instead, we modify the radio's positions under coarse localization and  rely on communications among them using Guided DBF.

\section{Guided Distributed Beamforming}
We start by explaining the concept behind Guided DBF, then we analyze its requirements in terms of  node positions, and the impact of positioning errors on the phases of the combining signals at the destination.
\begin{figure}[t!]
	\centering
	\includegraphics[width=3.3in]{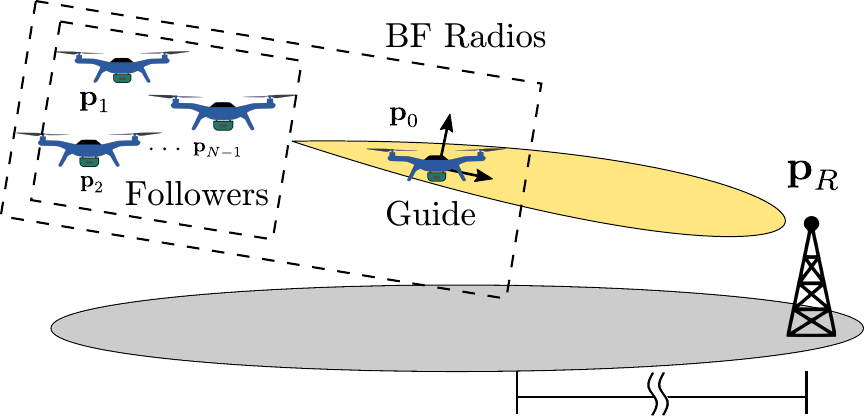}
	\caption{In the proposed approach, the followers beamform towards the guide, which repositions itself to lie on the line connecting the center of the cluster of followers to the destination.}
	\label{fig:proposed_approach}
\end{figure}
\subsection{Approach}
The proposed approach consists of having one of the DBF radios act as a  guide to the remaining radios, which are referred to as the followers.  The followers, using feedback from the guide, adjust their phases for coherent combining at the guide. %
By leveraging the radio's mobility,  the guide moves to be on the line  originating from the centroid of the followers towards the desired beamforming direction, making  the beamformed signals have a large combining gain at the destination receiver as shown in Fig.~\ref{fig:proposed_approach}. It is easy to see that if the guide was placed in the close vicinity of the destination, coherent combining at the guide would imply a large combining gain at the destination. However,  having one of the  beamforming nodes move near the destination defeats the purpose of beamforming. We want to attain the beamforming gain at the destination receiver without having any of the nodes travel a large distance. To that end, we study the relation between the positions of the nodes and the combining gains.  %

Without loss of generality, we assume that the node $0$ acts as the guide and that the followers are  nodes $1$ to $N-1$. To use Guided DBF, the followers need to cluster around the x-axis  ($\sum_{i=1}^{N-1} \mpy{i}\approx 0,\sum_{i=1}^{N-1} \mpz{i}\approx 0$) with $\mpx{i}\leq 0$ for $i \in \{1\cdots N-1\}$, making the line between the cluster of followers and the guide (which is the coordinate reference) point towards the DBF direction. If a LOS channel exists with the destination, the guide would lie on the line in between the followers and destination $\mpy{R}=0,\mpz{R}=0$ with $\mpx{R}>>0$.  

Since the followers beamform towards the guide, they adjust their phase based on the guide, that is they set their phases to $\theta_i=-\mhpg{i}$ where $\mhpg{i}$ is the phase of the channel between the guide and follower $i$ and is obtained using the guide's feedback. 
As for  the guide, it sends its signal without phase compensation, i.e, $\mwp{0}=0$, assuming it is hardware calibrated for phase reciprocity~\cite{guillaud_practical_2005}. Phase reciprocity implies that both its transmit and receive chains are phase calibrated to leverage channel reciprocity. 

Note that Guided DBF relies on the guide radio moving to change the direction of the beam, a process which can be slow. Hence, it is more suited to applications with one fixed destination  (like the ground station of search and rescue UAVs) than those with multiple destinations.
As stated earlier Guided DBF only requires knowing the direction to point the beam and does not require a LOS channel with the destination nor knowing its exact location.
 However, using these assumptions, it easier to determine the DBF direction, since it matches the LOS direction to the destination. That is why we assumed a LOS channel, modeled as a Ricean channel, between the DBF radios and the destination. This is the case if the destination radio is mounted on a high tower and the DBF radios are ground based vehicles in a rural area or UAVs. 
 
 Since the followers are adjusting their signals based on the guide and not the destination, the combining signals will have a phase mismatch  at the destination. This phase mismatch will lead to degraded BF gains. There two causes for this mismatch; first, the random non-LOS channel components  between the DBF radios and the destination.  These random NLOS components cannot be estimated and compensated  without the destination feedback. The second one is due to the deviation of the DBF radios from a line formation.  To limit the phase mismatch caused by the deviation from a line,  we want to determine the necessary separation between the followers and the guide to prevent the degradation of the BF gains.

\subsection{Guide Separation}
\begin{figure}[t!]
	\centering
	\includegraphics[width=3.3in]{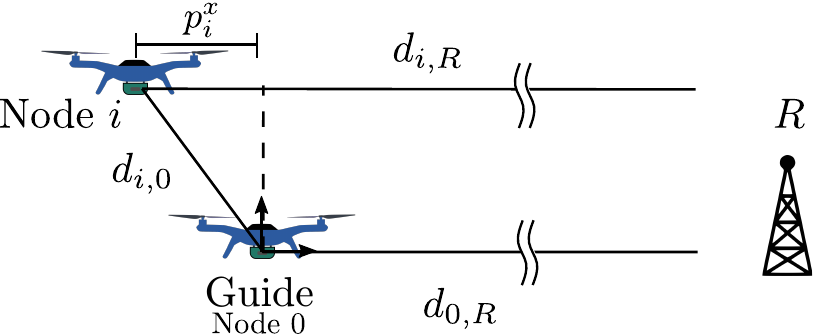}
	\caption{Assuming a distant destination, the signals to the destination follow almost parallel paths making the resulting phase error due to using the guide feedback   proportional to $ d_{i,0} - \mpx{i}$.}
	\label{fig:optimal_sep}
\end{figure}
To calculate the required guide separation, we consider the normalized expected value of the  channel, which is the geometric LOS component ($h_i=h_i^L= e^{j \frac{2 \pi d_{i,R}}{\lambda}}$). Since the NLOS channel components are random and unknown to the DBF radios, the radios cannot compensate for them. In that case, we can use the geometry
 to analyze the phase mismatch caused by using the guide for feedback instead of the destination. We start  by considering only a single  follower node $i$,  the guide, and destination as shown in Fig~\ref{fig:optimal_sep}. In Guided DBF, node $i$ adjusts its signal to have the same phase as the guide. So if the signal from node $i$ propagates through a distance $d_{i,0}+d_{0,R}$ it will be in phase with the signal transmitted from the guide.
   However, the signal for node $i$ propagates through a distance $d_{i,R}$ and not $d_{i,0}+d_{0,R}$. This makes the phase error between the guide  and node $i$  proportional to $d_{i,0}+d_{0,R} - d_{i,R}$. Assuming that the destination is distant, the signals from nodes $i$ and $0$ are almost parallel as shown in Fig~\ref{fig:optimal_sep}. In that case, the  mismatch in the propagation paths $e_{i}$ between the guide and node $i$
   \begin{equation}
   	\begin{aligned}
   		e_{i}&= d_{i,0}+d_{0,R}-d_{i,R}\\
   		& \approx  d_{i,0} - \mpx{i}
   	\end{aligned}
   \end{equation}
This mismatch of the different propagation paths will be translated to a phase error between the signals from the guide and radio $i$ given by
\begin{equation}
	\phi_{i} = \frac{2\pi e_i}{\lambda}
\end{equation}
So, by placing the nodes on a perfect line pointing toward the destination, beamforming towards the guide would guarantee coherent combining at the destination.
However, in practice due to the imperfect positioning caused to inaccurate localization or other mobility disturbances, a perfect line might not be practical to achieve.

\begin{figure}[t!]
	\centering
	\includegraphics{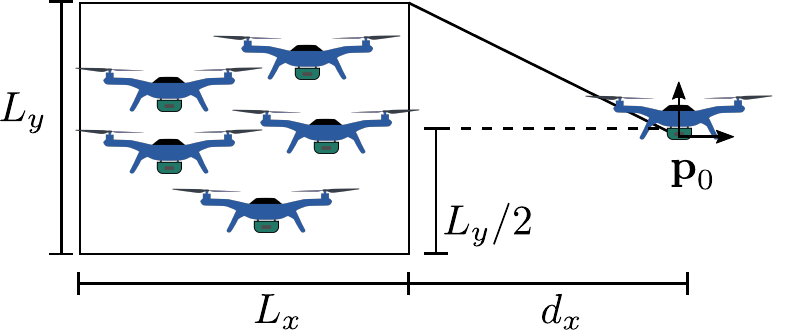}
	\caption{The followers locations are contained in a rectangle of dimensions $L_x$ and $L_y$, which is symmetric around the x-axis. The separation between the guide and the followers is given by $d_x$.}
	\label{fig:optimal_sep_all}
\end{figure}

After deriving the phase error for a single follower, we generalize it for all the followers. For simplicity, we limit our analysis to the 2D case ($\mpz{R}=0,\mpz{i}=0$ for all $i$) although the geometry can easily be extended to the 3D case. Hence, we define a rectangle of dimensions $L_x = \max_{i,j} |\mpx{i}-\mpx{j}|$ and $L_y=2 \max_{i} |\mpy{i}|$ having the vertical line $y=0$ at its center, which contains all the followers. The distance between the guide and the closest receiver in the x-dimension is given by $d_x=\min_{i} |\mpx{i}|$. This is shown in Fig.~\ref{fig:optimal_sep_all}. The upper bound of the propagation path mismatch due to using the guide, given by $e_{\max}$ is equal to 
\begin{subequations}
\begin{align}
e_{\max} &= \underset{i}{\max}\  e_i \\
& = \underset{i}{\max}\ \sqrt{ (\mpx{i})^2  + (\mpy{i})^2 }- \mpx{i} \\
& \leq \underset{i}{\max} \ \underset{j}{\max}\ \sqrt{ (\mpx{i})^2  + (\mpy{j})^2 }- \mpx{i} \label{eq:ub1} \\
&  = \underset{i}{\max}\ \sqrt{ (\mpx{i})^2  + (L_y/2)^2 }- \mpx{i}  \label{eq:ub11}  \\
&  \leq  \sqrt{ d_x^2  + (L_y/2)^2 }- d_x  \label{eq:ub2}
\end{align}  
\end{subequations}
where (\ref{eq:ub1}) adds another variable and can not decrease the maximization objective, and   (\ref{eq:ub2})  uses the fact that the function $\sqrt{a+x^2}-x$ is a strictly decreasing function in $x$ for any positive $a$ and $x$, and that $d_x\leq |\mpx{i}|$ for all $i$ by definition. This makes the largest phase deviation from the guide equal to $\phi_{\max}=\frac{2\pi e_{\max}}{\lambda}$. The smaller  $\phi_{\max}$, the larger the BF gains at the destination. The exact value of the DBF  gain ($\Gamma$)  depends on the  placements of the followers and is later considered in simulations.  %

Based on the tolerable amount of phase errors, we want to upper  bound  the path mismatch $e_{\max}$ by a chosen value of $\delta$  such that 
\begin{equation}
e_{\max} \leq \delta \label{eq:ub_val}
\end{equation}
The smaller the value of $\delta$, the smaller $\phi_{\max}$, which means less phase mismatch and   larger BF gains. Note that the chosen $\delta$ has to be   less than $\lambda$ for the maximum phase error $\phi_{\max}$ to be less than $2\pi$.  %
By manipulating~(\ref{eq:ub2}),  the relation between the guide separation $d_x$ and  vertical spread of the follower $L_y$ to realize (\ref{eq:ub_val}) for a given $\delta$ is 
\begin{equation}
d_x\geq\frac{(L_y/2)^2-\delta^2}{2\delta} \label{eq:opt_sep}
\end{equation} 
Hence, the separation between the guide and the followers ($d_x$) to achieve a given path mismatch $\delta$ scales quadratically with the vertical spread of the followers ($L_y$).  For a given follower placement (fixed $L_y$), using a smaller $\delta$ to reduce the phase errors requires the guide  to travel further to increase its separation. Hence, the choice of  $\delta$  trades off between the distance traveled by the guide and the BF gains as we will illustrate using simulations.

\section{Evaluating Guided DBF}
Using numerical simulations, for Guided DBF, we study the impact of the arrangement of the followers and the separation of the guide on the beamforming gain. Then we compare Guided DBF with Location DBF under localization errors. The impact of non-LOS channel components on Guided DBF is also evaluated.

In our simulations, we consider $N=11$ beamforming radios (1 guide and 10 followers). The follower nodes are assumed to be randomly placed in a rectangle of dimensions $L_x \times L_y$ at a distance $d_x$ from the guide as shown in Fig.~\ref{fig:optimal_sep_all}.   The receiver $R$  is placed at a distance of 10KM from the guide. The frequency used in the simulation is 900MHz making the wavelength equal to $33.3$cm. The considered channel is Ricean with a K-factor of 25dB, a value typical of air-to-ground channels in near-urban or suburban regions~\cite{matolak_airground_2017}.   The channel estimates between the guide and the followers are assumed to be perfect, making the combining gain at the guide always equal to 1. This assumption is justified later in  Section~\ref{sec:comparison}, where we simulate the signals exchanged for synchronization. The beamforming gain in our results is the gain of all the DBF radios as measured by the destination receiver.

\begin{figure}[t!]
	\centering
	\includegraphics{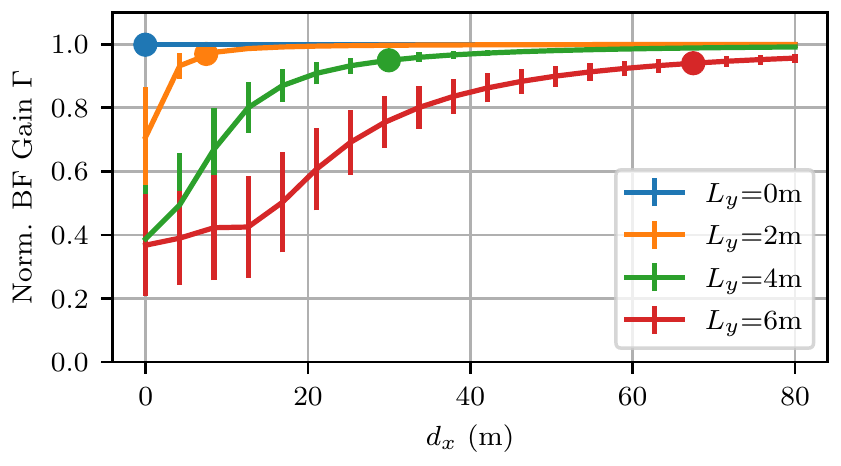}
	\caption{The BF gain of random placement of the followers for $L_x=10m$ for different values of $L_y$ as a function of the separation of the guide $d_x$.}
	\label{fig:orientation_array}
\end{figure}

\subsection{Impact of DBF Nodes  Geometry}
First,  we consider the effect of the separation between the guide and the followers $d_x$ on the combining gain at the destination receiver $\Gamma$. This evaluation is performed for $L_x=10m$ and for different $L_y$ as shown in Fig.~\ref{fig:orientation_array}. For each point, we consider 100 uniform random placements of the followers within the deployment rectangle and plot the mean BF gain with error bars representing the standard deviation.  We can see that as  $L_y$ increases the guide needs to be further from the followers to ensure coherent combining at the receiver.  In the case of linear array, $L_y=0$, the optimal combining gain can be attained with no separation ($d_x=0$). In Fig.~\ref{fig:orientation_array}, the solid circles show the combining gain when using the optimal separation calculated using~(\ref{eq:opt_sep}) for a tolerable mismatch given by $\delta=0.2\lambda$. These circles show that using  (\ref{eq:opt_sep}) and for this choice of $\delta$, most of the BF gains are attained, despite of the Ricean channel non-LOS components for $K=$25dB. The impact of the Ricean channel K-factor is further studied  later. Also, the relation between $\delta$ and the BF gains is further discussed later. 

\begin{figure}[t!]
	\centering
	\subfloat[The beamforming gain for mismatch tolerance $\delta$ as a function of $L_y$.\label{fig:optimal_dist_gn}]{\includegraphics{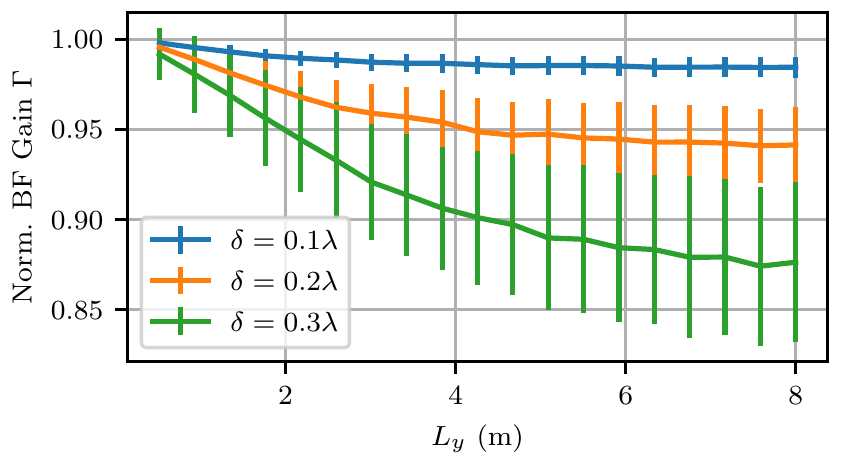}} \\
	\subfloat[ The separation of the guide obtained using (\ref{eq:opt_sep}) for different value of the mismatch tolerance $\delta$ as a function of $L_y$. \label{fig:optimal_dist}]{\includegraphics{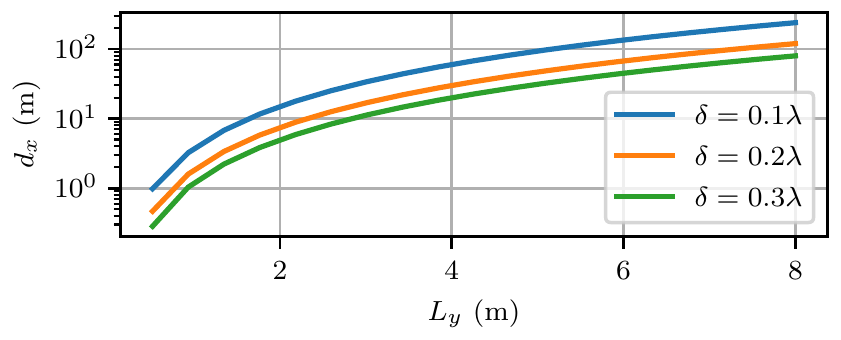}} 
	\caption{ The beamforming gain and the distance traveled for different values of mismatch tolerance $\delta$.}
	\label{fig:optimal_dist_all}
\end{figure}

To get an understanding of the separation between the guide and the followers as a function of the vertical spread of the followers ($L_y$), we plot the lower bound from (\ref{eq:opt_sep}) in Fig.~\ref{fig:optimal_dist} on a logarithmic scale. For our simulation setup, a vertical spread below 1m would require separation  below 2m. 
For larger spreads up to 8m, the separation can be over 100m. This is expected since (\ref{eq:opt_sep}) is quadratic in $L_y$.  
Hence, it is beneficial to align the followers to avoid large displacement of the guide. 

We also consider the effect of the chosen tolerance on the distance traveled and the combining gain. 
As predicted by (\ref{eq:opt_sep}), a tighter tolerance requires larger separation. In terms of combining gain at the end receiver, in accordance with the result in~\cite{mudumbai_feasibility_2007}, the combining gain is tolerant to phase errors which are due to mismatch between the guide and the destination. 
Fig.~\ref{fig:optimal_dist_gn} shows that a tolerance of $0.2\lambda$ is able to attain over 90\% of the combining gain while requiring at about half the separation of $0.1\lambda$ as shown in Fig.~\ref{fig:optimal_dist}. Note that the BF gains change for the same $\delta$ because  $\delta$  only imposes an upper bound on the phase error. For the same $\delta$, since the radios are randomly placed within a rectangle, the distribution of the phases vary with $L_y$ and $d_x$ leading to varying BF gains. 

\begin{figure}[t!]
	\centering
	\includegraphics[scale=0.8]{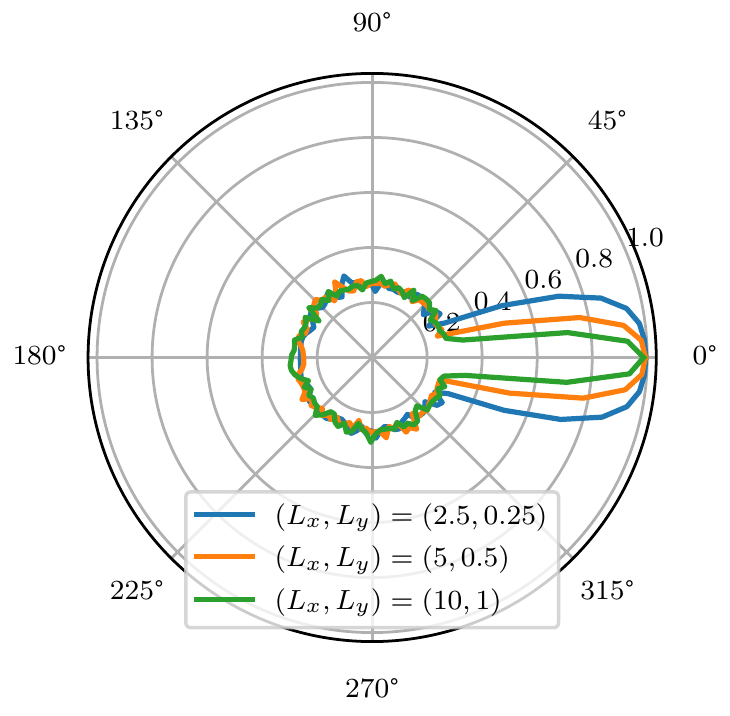}
	\caption{The beamforming pattern for different deployment regions for $N=11$. The values of $L_x$  and $L_y$ determine the beamwidth.}
	\label{fig:beampattern}
\end{figure}
Then we simulate the beampattern  obtained in the far field when the followers beamform toward the guide. In Fig.~\ref{fig:beampattern}, we plot the average  beampattern when the followers are deployed randomly for different regions  the guide placed at $d_x$ to achieve $\delta=0.2\lambda$ using~(\ref{eq:opt_sep}). For $L_x=10$m and $L_y=1$m, we can observe that the realized beampattern has a narrow beamwidth. As the region becomes smaller, the beamwidth becomes larger. Thus by changing the deployment region, we can control the beamwidth. %

\subsection{Localization Errors }
We evaluate the sensitivity of our proposed approach to localization errors of the DBF radios and compare it against  Location DBF, where only locations are used calculate the beamforming weights. For Location DBF, we assume that the carriers of  all radios are  phase synchronized. The localization errors are modeled as uniform random variables $\Delta p^x_i$ and $\Delta p^y_i$ which are unknown to the nodes. These errors take values between $-\Delta P /2$ and $\Delta P /2$ added to the positions of the nodes where $\Delta P$ is the error range, such that node $i$ would be located in $[\mpx{i} + \Delta p^x_i, \mpy{i} + \Delta p^y_i]$ for $i\in \{0,1,\cdots,N-1\}$ (the origin is assumed to be fixed regardless of  localization errors). For location based beamforming, we set the beamforming of node $i$ to $\mwp{i} = e^{ -j\frac{2\pi}{\lambda} d_{i}}$, where $d_i = \mpx{i} + L_x$. 
This choice of $d_i$ only uses the known position $\mpx{i}$ and  ensures that the resultant wave from the radios  is aligned pointing towards the receiver in the case of no localization errors. We consider the same initial setup with $L_x=10$ and $L_y=1$.
For the Guided DBF, we accounted for the worst case localization error by  increasing the separation $d_x$ using $L_y=1+ \Delta P$  in (\ref{eq:opt_sep}) for $\delta=0.2 \lambda$.
We consider both the case where the guide suffers from localization error similar to the rest of the nodes and a perfect guide which does not suffer from localization errors.

\begin{figure}[t!]
	\centering
	\subfloat[The effect of localization errors on Guided DBF, Guided DBF assuming an the guide perfectly placed (Perf. Guide), and using Location DBF.\label{fig:location_based_gn}]{\includegraphics{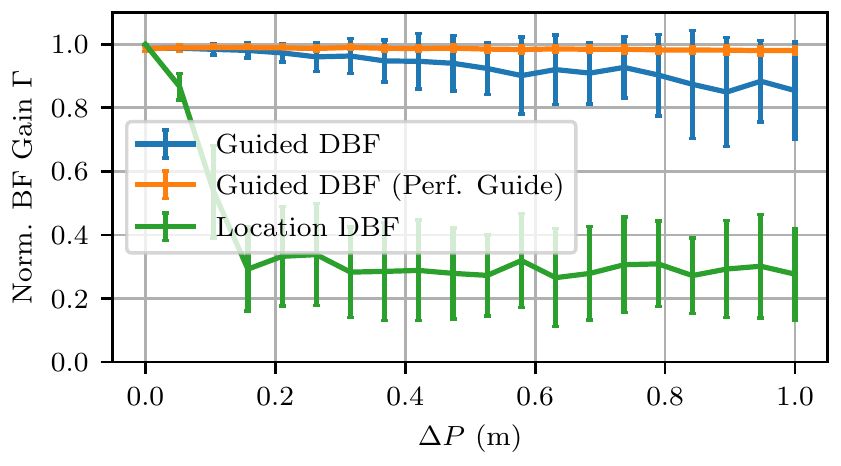}} \\
	\subfloat[The sepration of the guide as a function of the range of localization error $\Delta P$.\label{fig:location_based_dist}]{\includegraphics{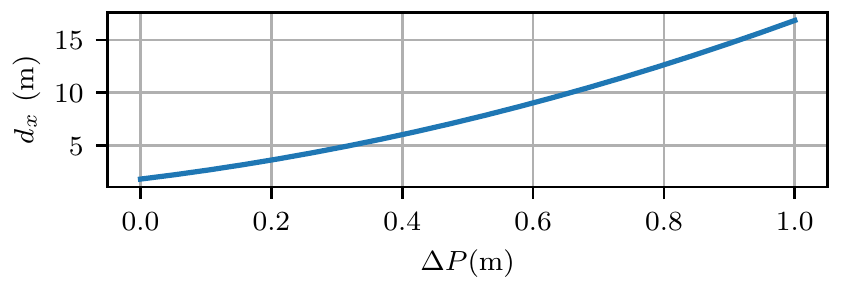}} 
	\caption{The effect of localization error on the guide based beamforming and location based beamforming. The wavelength used is $\lambda=0.33$m}
	\label{fig:location_based}
\end{figure}
The combining gain obtained only using location information is shown in Fig.~\ref{fig:location_based_gn} against the localization error range $\Delta P$ for $L_x=10$ and $L_y=1$. We can see that for perfect location information, a gain of 1 is attained using Location DBF but as localization error range increase, the beamforming gain decreases rapidly and becomes as good as random when the magnitude of localization error approaches $\lambda/2$ (16.6cm). This shows that Location DBF requires  location information accurate within a fraction of a wavelength to work. While localization systems that can attain this accuracy exist, they require a large bandwidth~\cite{khelifi_survey_2019}, which is not always available. %
  Guided DBF, on the other hand, is maintaining average BF gains above $0.8$ even as the localization error reach $1$m, which is equivalent to $3\lambda$. However, it still decays to some extent. 
This decay is explained by the localization errors in the guide making the beam formed by the followers point slightly towards the wrong direction. Since the beam for $L_x=1$m and $L_y=10$m is narrow as we have shown in Fig.~\ref{fig:beampattern}, this  leads to suboptimal combining gains. 
However, this can be resolved by choosing the placement to attain a wider beam (using smaller $L_x$ and $L_y$). The used separation of the guide, which increases with the range of the error is shown in Fig.~\ref{fig:location_based_dist} and it does not exceed $18$m for our setup.  This shows that our proposed approach can compensate for localization errors by reasonably increasing the separation of the guide to account for the worst-case vertical spread.

\begin{figure}[t!]
	\centering
	\includegraphics{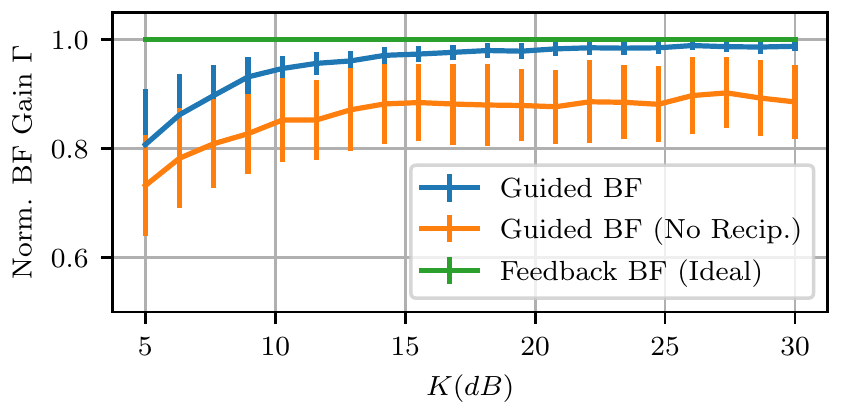}
	\caption{The effect of non-LOS channel components, simulated using a Ricean channel, on the BF gains of different DBF approaches.}
	\label{fig:ricean_chan}
\end{figure}

\subsection{Channel Induced Errors}
Previously, we only considered a Ricean channel with a 25dB K-factor. In this subsection, we study the performance of Guided DBF under different $K$-factor. The smaller $K$, the stronger the non-LOS channel components. The results,  obtained from the same simulation setup, are shown in Fig.~\ref{fig:ricean_chan}. From this Figure, we can see that the BF gains of Guided DBF degrade for smaller $K$. This is expected since the non-LOS channel component is unknown at the DBF radios and cannot be estimated without any feedback from the destination. Yet for LOS dominant channels with K-factors above 15dB, Guided DBF retains almost all the BF gains.  

Then, we consider the impact of a non-reciprocal guide. Depending on the RF front end implementation, the phase offset between the guide's transmit and receive chains can be  varying between transmissions making the guide phase non reciprocal.  When we consider a non reciprocal guide, simulated by making its phase uniformly random, the BF gains drop by a approximately $1/N$ as shown in Fig.~\ref{fig:ricean_chan}. This happens because  the signal transmitted by the guide is not necessarily coherently combining with the signals of the followers at the destination. This incoherent combining is because the followers are adjusting their signals based on the guide's receive chain, which has a different phase from its transmit chain. 

From the same Figure~\ref{fig:ricean_chan}, as expected,  we see that  DBF using ideal feedback is not affected  by the non-LOS components as they are estimated and compensated for. However, in practice, for a far destination, the feedback is not ideal and has errors which make the BF gains degrade significantly.  In order, to show the impact of feedback error, we need to simulate the signaling between the destination and the DBF radios according to the DBF protocol. While Guided DBF would work using any DBF protcol, yielding coherent combining at the guide,  in  Section~\ref{sec:implementation}, we describe a specific DBF protocol. This protocol is used for the DBF protocol comparison in Section~\ref{sec:comparison} and later in the experimental proof of concept in Sec.~\ref{sec:experiment}.

\section{Distributed Beamforming Protocol}
\label{sec:implementation}
\begin{figure}[t!]
	\centering
	\includegraphics[scale=0.85]{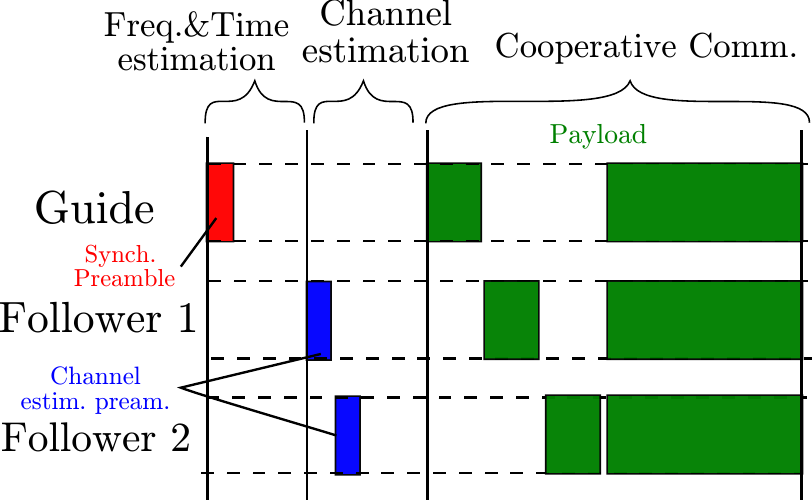}
	\caption{The timing diagram of the implemented beamforming setup for $N=3$. The guide initiates the beamforming by sending a synch. preamble. The followers then, after performing time and frequency corrections, send channel estimation preambles. The payload shown in green was designed such that each DBF radio transmits individually, then all of them beamform, to enable the evaluation of the beamforming gain.}
	\label{fig:exp_timing_diagram}
\end{figure}
 In this section, we describe the DBF protocol used in the comparison of DBF approaches  and the experiments to achieve coherent combining at  a radio providing feedback. For Feedback DBF, this radio is the destination and for Guided DBF it is one of the DBF radios, which is the guide. Note that the Guided DBF would work using any DBF protocol as long as the signals from the followers combine coherently at the guide. The used DBF protocol  was chosen to use explicit channel feedback as it is more robust to channel variations than iterative approaches like 1-bit feedback~\cite{hanna_distributed_2021}.

 As part of the DBF protocol, in addition to adjusting their phases, since each DBF radio has its own local oscillator and timing clock, the radios need to first synchronize in  frequency to avoid phase drift and in time to avoid intersymbol interference. 
To  achieve these requirements each node $i$  estimates its frequency offset $\mdf{i}$ and timing offset $\mdt{i}$, relative to the radio providin feedback along with the channel phase estimate $\mhp{i}$. After estimating and digitally correcting for these errors, coherent combining can be attained at the feedback radio. 

A protocol to achieve coherent combining was developed for software-defined-radios (SDR) having a  sampling time  $T_s$. A timing diagram of the protocol  is shown in Fig.~\ref{fig:exp_timing_diagram}.  It consists of a frequency and timing estimation stage which aims to estimate $\mdf{i}$ and  $\mdt{i}$, followed by a channel estimation stage to obtain $\mhp{i}$. Afterward, the radios transmit their payload.

The beamforming is initiated when the guide transmits a synchronization preamble. The timing and frequency estimation is performed simultaneously using this preamble using the approach from~\cite{yan_aeroconf_2019} as follows: The DBF radios obtain a one-shot frequency estimate from this preamble and apply the extended Kalman filter (EKF)  as an averaging filter~\cite{quitin_scalable_2013}. For the timing estimation, the time of arrival (TOA) of the preamble is used as a reference for timing~\cite{elson_fine-grained_2003}. The TOA is estimated by using correlation for sample level timing accuracy and maximum likelihood is used for sub-sample-time accuracy~\cite{yan_aeroconf_2019}.

After frequency and timing estimation, the channel is estimated. We use explicit channel estimation, where each of the followers is assigned a time slot to transmit a preamble. The guide estimates the phase of the received preamble and feeds it back to each follower either in-band or through a side-channel. At the last stage, radio $i$ transmits its payload after correcting for timing and frequency offsets and using $\mwp{i}=-\mph{i}$. In practice, there are errors in channel estimation, frequency and timing synchronization, which lead to imperfect combining gains at the guide despite of having feedback. These errors increase with as the feedback radio is at a further distance from the DBF radios.

In both the signal-level simulation and SDR implementation, we considered a sampling rate of 1Mbps equivalent to $T_s=1\mu s$, a synchronization preamble of  630$\mu s$ duration and the channel estimation preamble  of 200$\mu s$. The first stage was allocated 60ms, the second stage 20ms, and the third stage 30ms.  The time assigned for each stage contains guard times  for the real-time processing when the protocol is implemented using SDRs as discussed later. 

\section{Comparison of Beamforming Approaches }
\label{sec:comparison}
\begin{figure}[t!]
	\centering
	\includegraphics[scale=1]{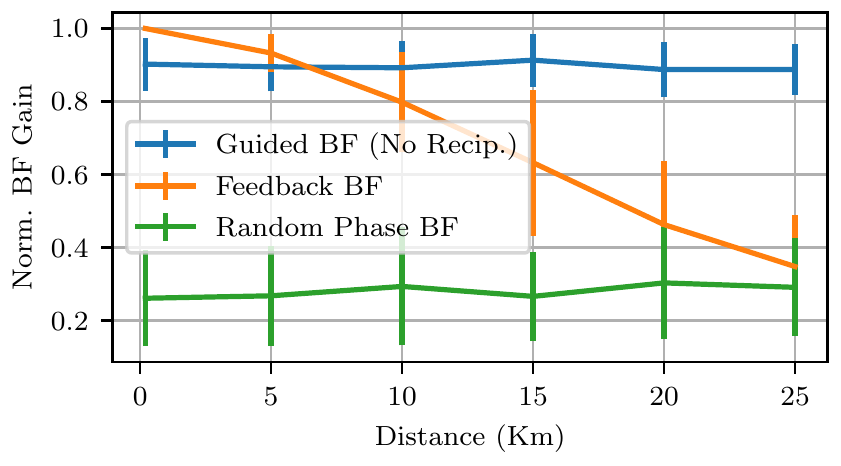}
	\caption{Comparison between Guided DBF, Feedback DBF, and Random Phase DBF as a function of the destination distance using signal level simulation. A Ricean channel with a 2.3 channel exponent path loss is simulated. }
	\label{fig:exp_comparison}
\end{figure}
In this section,  we compare Guided DBF with Feedback DBF and Random DBF considering the path loss, which increases with the average distance to the destination ($d=\frac{1}{N} \sum_i d_{i,R}$). The same environment is used in simulations with $N=11$,  $L_x=10$ and $L_y=1$. We consider  a Ricean channel having a path loss with  a  2.3  channel exponent. The considered channel K-factor decreases with distance (stronger NLOS), as obtained from air-to-ground measurements  in near-urban  environment~\cite{matolak_airground_2017}. To be specific, we considered $K=K_0 - \eta_0 (d-d_0) + Y$, where $K_0=29.9$, $\eta=0.02$ $d_0=3.4$, and $Y$ is a zero mean Gaussian random variable of 2.2 standard deviation.  We consider DBF radios having a transmit power of $P_t=0$dBm,  a noise bandwidth of 1MHz and a 5dB noise figure. Using this bandwidth, for an air-to-ground channel having a 11 ns median root-mean-square delay spread~\cite{matolak_airground_2017}, the  narrowband channel assumption is justified. Using this setup, with a perfect BF gain, the SNR at the destination exceeding 1dB can be attained at 25Km.  For  Feedback DBF, we simulated the previously described protocol between the destination and the DBF radios; we generated each signal at the SNR  according to the  distance and estimated the frequency offset using EKF and channel estimation as described in~\cite{yan_aeroconf_2019}. The destination radio was assumed to have 20dBm transmit power, i.e, 100 times more power than the DBF radios for Feedback DBF. For the Guided DBF, we assumed that the guide is not reciprocal. As for the Random  DBF, it was simulated by using beamforming weights with uniformly distributed random phase. 

The results are shown in Fig.~\ref{fig:exp_comparison} as a function of the distance from the destination ($d$).   From that Figure, we can see that for  $d$ smaller than 6Km, Feedback DBF provides larger BF gains than Guided DBF because, unlike Guided DBF, it is not affected by the non-LOS channel components, nor by the non-reciprocity of the guide. However, as the distance increases beyond 6 Km, the SNR of the feedback signals exchanged with the destination becomes low,  even though the destination has 100 times the power of the DBF radios. This degradation is caused by synchronization errors and channel estimation errors which make Feedback DBF start to behave like  Random DBF at 25 Km. Using Guided DBF, on the other hand, the DBF feedback is from the guide, placed at a few meters away from the followers, regardless of the distance from the destination. Hence its feedback is always at high SNR. Even though, Guided DBF is affected by the K-factor which increases with the distance, it still outperforms  Feedback DBF at a large distance. The BF gain of Random  DBF, on the other hand, is not affected by distance, since it also does not rely on destination feedback. However, its BF gains are much lower than the other two approaches. So, the same signals needs to be transmitted multiple times so that it combines coherently at least once. Note that for Random BF, the fewer the number of radios the more likely they will combine coherently as we will see later in our experimental proof of concept.
\begin{figure*}[t!]
	\centering
	\includegraphics{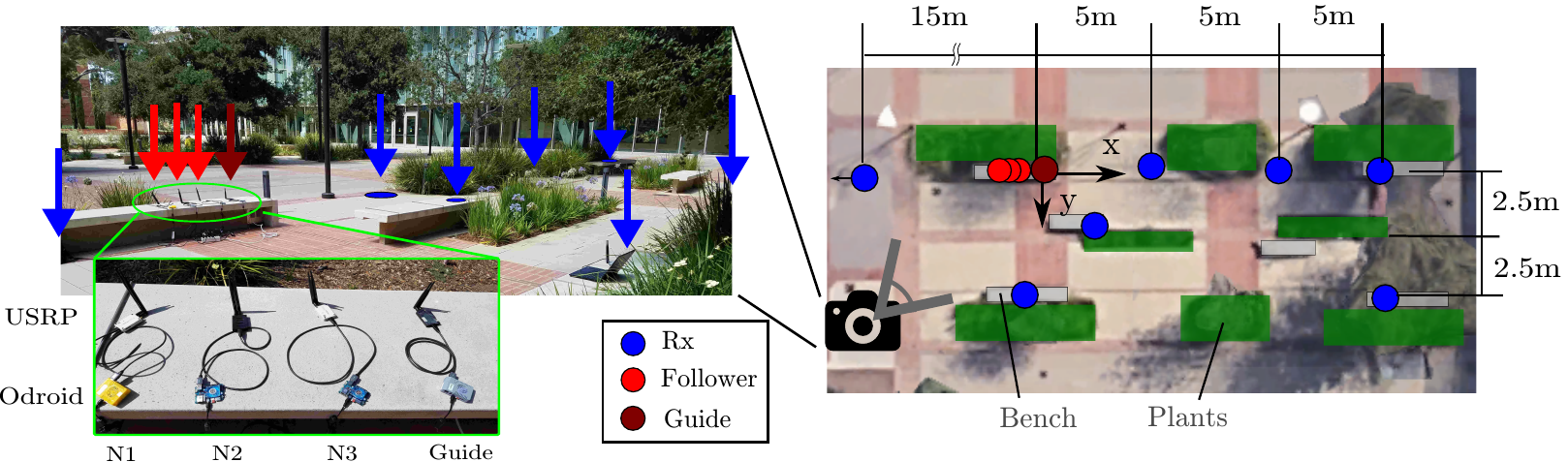}
	\caption{For the experimental proof of concept, the 4 DBF radios were placed coarsely in a linear arrangement in an outdoor environment. The destination receiver was moved in the places highlighted in blue.\label{fig:exp_env}}
\end{figure*}

Although the results show that Guided DBF provides larger BF gains than Random DBF and Feedback DBF at large distances in a LOS channel, it is important to note that Random DBF and Feedback DBF have their advantages making them more suitable for other deployment scenarios.  Guided DBF requires a LOS dominant channel to work and requires placing the DBF radios in a specific geometric arrangement, making it suitable for mobile radios in open areas like UAVs deployed over suburban areas. Feedback DBF would work in any channel as long as the destination feedback is at an adequate SNR. Random DBF, while it sacrifices throughput, it is simple to implement as no feedback nor communication between DBF radios are needed. Table~\ref{tbl:comparison} compares the underlying assumptions of the  different DBF approaches along with Location DBF.  Guided DBF can provide coherent combining using coarse placement, without requiring phase synchronization between the DBF radios, unlike Location DBF. It does not require any destination feedback like Feedback DBF, nor requires multiple repetitions of the same data like Random DBF. In our simulations, we implicitly assumed that the radios are isotropic point sources, which is not the case in practice. To validate Guided DBF, we consider an experimental proof of concept using SDRs.

\begin{table}[t]
	\renewcommand{\arraystretch}{1.4}
	\caption{Comparing between DBF Approaches \label{tbl:comparison}}
	\centering
	\begin{tabular}{|p{0.8in}|c|c|c|c|}
			\hline
			Assumption & Location  & Feedback   & Random  & Guided \\	\hline
			LOS Channel & No  & No  & No & Yes\\	\hline
			Geometric Arrangement & No & No  & No  &Yes \\	\hline
			Localization & \textbf{Accurate} & No  & No  & \textbf{Coarse} \\	\hline
			InterRadio Phase Synch & \textbf{Yes} & No  & No  & \textbf{No} \\	\hline
			Dest. Feedback & No & \textbf{Yes}   & No & \textbf{No} \\	\hline
			Multiple Transmissions & No & No  & \textbf{Yes}  & \textbf{No}  \\	\hline
		\end{tabular} 
\end{table}

\section{Experimental Proof of Concept}
\label{sec:experiment}
Guided DBF is envisioned for  communications over distances in the range of kilometers, for practical reasons, our experimental setup has at a much smaller scale of tens of meters. At this small scale,  we aim to only provide a proof of concept of  Guided beamforming; we want to verify that Guided DBF can provide significant BF gains along the desired direction in a realistic  LOS channel using real radios. Also, we want to gain insights into the generated beampattern of Guided beamforming and whether it can be used with coarsely placed radios and uncertain destination location.  

\begin{figure}[t!]
	\centering
	\includegraphics[scale=0.8]{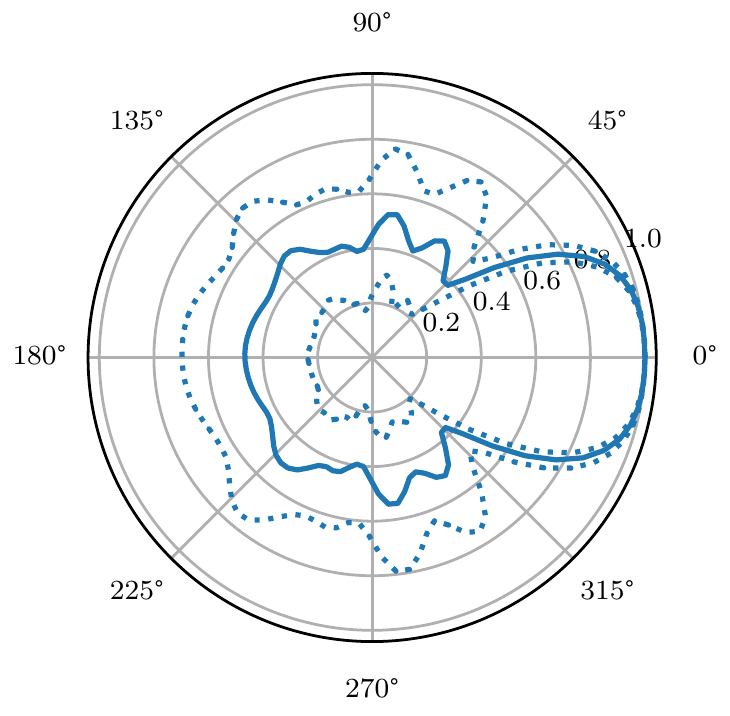}
	\caption{The beamforming pattern for $L_x=0.55$, $L_y=0.1$  emulating the experimental setup. The solid line is the average of 100 random placements and the dotted lines represent $\pm$ their standard deviation. \label{fig:beampattern_exp}}
\end{figure}

\begin{figure*}[t!]
	\centering
	\includegraphics{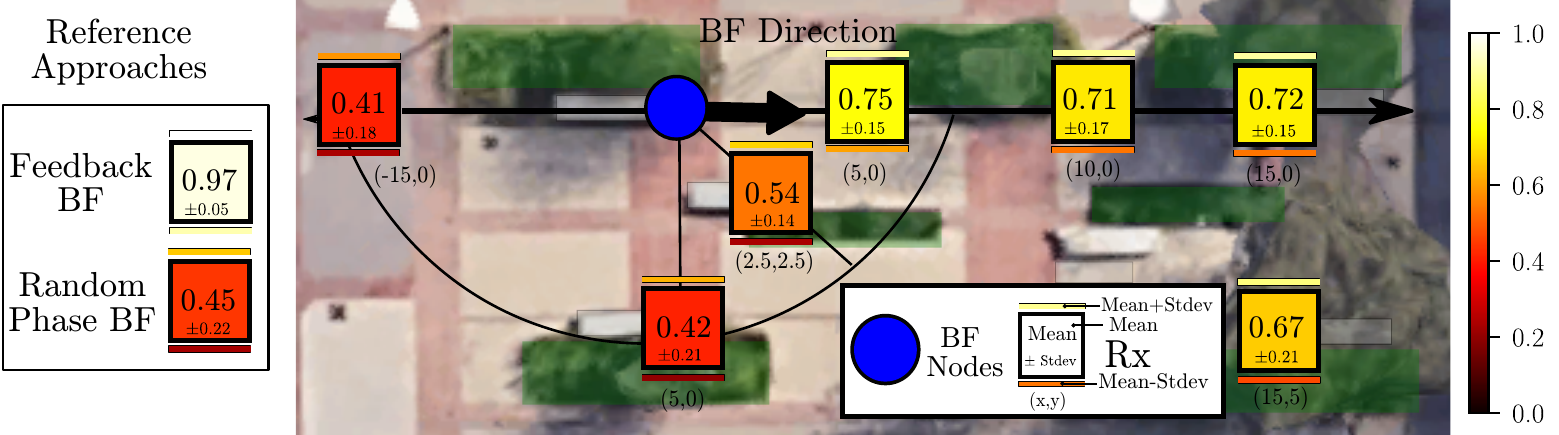}
	\caption{The results from the experimental evaluation are shown over the top view of the environment. The mean BF gain at each destination receiver (Rx) location is written inside a square colored according to the heatmap along with the standard deviation (Stdev). Stripes above and below each square show  the mean plus and minus the Stdev respectively.}
	\label{fig:exp_results}
\end{figure*}
\begin{figure}[t!]
	\centering
	\includegraphics{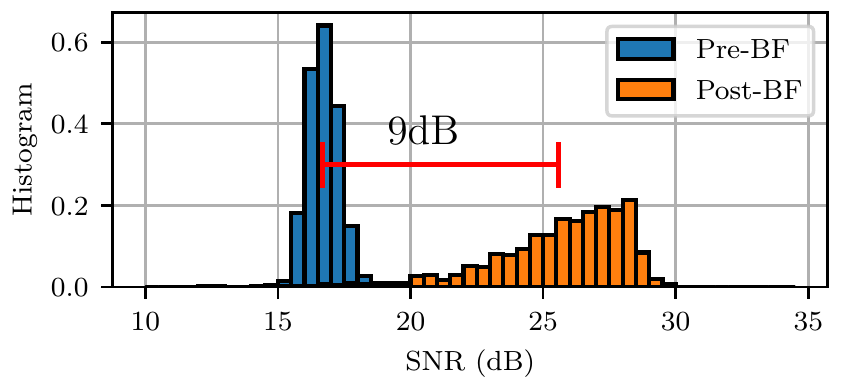}
	\caption{The histogram of pre-DBF and post-DBF SNR at (15,0). A 9dB mean SNR improvement was obtained by DBF.}
	\label{fig:exp_snr}
\end{figure}
\subsection{Implementation and Evaluation Procedure}
The protocol was implemented using GNU Radio~\cite{noauthor_gnu_nodate}  and  the USRP B205-mini software defined radio~\cite{ettus_research_usrp_nodate}. The center frequency used is  915MHz and 
the timing of the different transmissions within the packets was ensured by using the USRP hardware driver (UHD) timing tags. The channel feedback was implemented using a WiFi side channel.

As for the hardware, single board computers (SBC), namely the ODROID XU4~\cite{hardkernel_odroid-xu4_nodate}  were used to power the guide and the followers. The end receiver was operated using a laptop.  Note that since the DBF protocol needs to run in real time, the waveforms were designed to work efficiently using SDRs powered by computationally constrained SBC and do not follow any standardized communication protocol.

While the entire payload can be used for beamforming, the payload was designed such that each of the followers transmits individually in a portion of the time, and a portion assigned for beamforming as illustrated in Fig.~\ref{fig:exp_timing_diagram}.  To evaluate the beamforming gain of a single packet,   the magnitude of the beamformed signal and that of the sum of the individual transmissions of the nodes are substituted in~(\ref{eq:DBF_gain}). 

The results consist of several measurements. A single measurement consists of 900 packets captured over a period of 5 minutes. To evaluate the BF protocol, the payload is designed such that each node transmits individually, then they beamform  as illustrated in Fig.~\ref{fig:exp_timing_diagram}. The beamforming gain of a single packet is calculated using  the magnitude of the beamformed signal and that of the sum of the individual transmissions of the nodes according to~(\ref{eq:DBF_gain}). 
The individual transmission of one node consists of 4K samples chosen to be all ones, hence yielding an unmodulated carrier. The beamforming portion consists of 14K samples, out of which 4K samples are ones and the remaining 10K samples are BPSK modulated  using root raised cosine pulse shaping with 2 samples per symbols. This makes  the signal bandwidth at 1Msps equal to 500KHz. %

\subsection{Experimental Setup}
The Guided beamforming experiment was performed in an outdoor environment as shown in Fig.~\ref{fig:exp_env} as a top view and a camera photo. Five  nodes were used in the experiment, three followers, one guide, and one node as the destination. The four beamforming nodes were all placed in a coarse linear arrangement as shown in Fig.~\ref{fig:exp_env} making  the beamforming direction is towards the positive x-axis. This arrangement can be described according to our system model as having $L_x=0.55$m, $L_y=0.1$m. The guide separation was set to $d_x=0.32$m, which exceeds the value calculated using (\ref{eq:opt_sep}) for $\delta=0.1 \lambda$. To evaluate the beamforming gains, the destination receiver was moved to several locations   as shown in blue in the same Figure. 

To obtain an expectation of the performance at different locations based on theory, we  simulated  the beampattern expected from the DBF node placement.  To do that we followed the same procedure of randomly placing $N=3$ nodes within $L_x$ and $L_y$. The results are shown in Fig.~\ref{fig:beampattern_exp} with the solid line representing the mean and the dotted lines for $\pm$ the standard deviation. From this Figure, we expect a  wide beamwidth in the BF direction. While the beampattern gives as an expectation based on theory, an experimental baseline is needed to interpret the Guided DBF gains.

As a baseline, we consider both Feedback DBF and Random DBF. Both approaches were evaluated  using a different experimental setup, which took place in a lab environment. In this setup,  the DBF radios were arranged in a line of length $0.7$m broadside to the destination placed at a distance of 2.5m, making the SNR of feedback from the destination very high.  The purpose of evaluating Feedback DBF and Random DBF is not making a direct comparison, which is not possible at a small scale experimental setup, but to place the Guided DBF results in context;
 Feedback DBF serves as an experimental upper bound for the performance of Guided DBF since the destination feedback is at a high SNR and hence is close to ideal.   As for Random DBF, it serves only as a baseline, which Guided DBF needs to significantly outperform. Note that since we are considering few BF radios ($N=4$), the normalized BF gains from Random BF will be much higher than the previous simulations for $N=11$.

\subsection{Results}
To demonstrate Guided DBF, we measured the Guided DBF gains for different destination locations. The results are shown in Fig.~\ref{fig:exp_results} imposed over the top view of the environment.  For each measurement location, the average BF gain is written inside a square whose color follows the heatmap shown on the left of the Figure. The standard deviation (Stdev) of the measurements is also written and shown as colored stripes above and below each square. The (x,y) coordinate of each measurement location is written beneath the square. Both the Feedback and Random DBF results are shown outside the top view. 

Looking at location (5,0), it has a large average BF gain 0.75, which is significantly higher than that of Random DBF (0.45). Yet, it is  lower than that of Feedback DBF (0.97) mainly because the Guide lacks phase reciprocity. Without  reciprocity,  the signal transmitted from the guide has an almost random phase which causes about 1/N (0.25) drop in BF gains. Another reason for this drop is that Guided DBF does not account for the random reflections in the environment. Looking at the other positions in the DBF direction --- (10,0), and (15,0) --- they also have  a large and steady BF gain of about 0.72.  This shows that the BF gains are consistent along the desired DBF direction despite the coarse placement of the radios.   To illustrate the impact of these BF gains, in Fig.~\ref{fig:exp_snr}, we show the histogram of the prebeamforming  SNR (calculated as the average for the 4 nodes) and postbeamforming SNR, for (15,0). From that Figure, we see that, the SNR improved by about $9$dB on the average.%
Since Guided DBF does not rely on any feedback from the destination, the BF gain are expected to hold for further distances and thus extend the range of communications. 
 
 Considering, the point (15,5) on the bottom left of  Figure~\ref{fig:exp_results}, we see that it also has  a relatively high BF gain of 0.67, which indicates a large beamwidth, which is consistent with the simulated beampattern. Since the beamwidth is large for this setup, we do not need accurate knowledge of the location of the destination to point the beam.
 From the same Figure, as we move away from the DBF direction along the points  (2.5,2.5), (0,5), and (-15,0), the BF gains degrade further as predicted by the simulated DBF pattern. This shows that beampattern obtained using Guided DBF behaves as predicted by simulations. Thus we can use simulations to design the deployment region of the DBF radios to obtain a desired pattern, taking into consideration how accurately we know the destination location.

\section{Conclusion}
We  proposed Guided distributed beamforming as an approach for mobile radios sharing a LOS channel to make their signals coherently combine at a remote destination,  without feedback from the destination and without requiring sub-wavelength location information.  We have derived and verified using simulations a geometric criterion for the placement of the radios. Unlike Location DBF, Guided DBF was shown to tolerate localization errors up to a few wavelengths. The BF gains from Guided DBF were shown to be independent from the distance to the destination, unlike Feedback BF, making it suitable for range extension.    Guided DBF was implemented using software defined radios. The results show a 9 dB  SNR improvement in the beamforming direction when using 4 radios and that the obtained measurements follow the beampattern predicted by simulations.

\ifCLASSOPTIONcaptionsoff
  \newpage
\fi

\bibliographystyle{IEEEtran}
\bibliography{references}

\end{document}